\documentstyle [12pt,aasms4]{article}
\input epsf

\journalid{}{}
\articleid{}{}

\begin{document}
\def\lax    {\ifmmode{_<\atop^{\sim}}\else{${_<\atop^{\sim}}$}\fi}
\def\gax    {\ifmmode{_>\atop^{\sim}}\else{${_>\atop^{\sim}}$}\fi}
\def\gtorder{\mathrel{\raise.3ex\hbox{$>$}\mkern-14mu
             \lower0.6ex\hbox{$\sim$}}}
\def\ltorder{\mathrel{\raise.3ex\hbox{$<$}\mkern-14mu
             \lower0.6ex\hbox{$\sim$}}}

\title{MASS DETERMINATION of BLACK HOLES in LMC X-1 AND NOVA MUSCAE 1991
 FROM THEIR HIGH-ENERGY SPECTRA}

\author{ C.R. Shrader\altaffilmark{1,2}
and Lev Titarchuk\altaffilmark{1,3} }

\altaffiltext{1}{Laboratory for High--Energy Astrophysics,
NASA Goddard Space Flight Center, Greenbelt, MD 20771, USA;
shrader@grossc.gsfc.nasa.gov,titarchuk@lheavx.gsfc.nasa.gov}
\altaffiltext{2}{Universities Space Research Association,Lanham MD}
\altaffiltext{3}{George Mason University/Institute for
Computational Sciences and Informatics, Fairfax VA}

\rm

\vspace{0.1in}

\begin{abstract}

We offer a brief description of the bulk-motion Comptonization (BMC) model for
accretion onto black holes, illustrated by its application
to observational data for LMC~X-1, and Nova Muscae 1991. We then extract some
physical parameters of these
systems from observables {\it within the context of the BMC model}, drawing from
results on GRO~J1655-40, for which we presented extensive
analysis previously. We derive estimates of the mass, ($16\pm 1
)M_{\sun}\times[0.5/\cos(i)]^{1/2}$ and mass accretion rate in the disk in
Eddington units ($\dot m_d\simeq 2$) for LMC X-1, and ($24\pm
1)M_{\sun}\times[0.5/\cos(i)]^{1/2}d_{5.5}$ and  ($\dot m_d\simeq 3$ ) for Nova
Muscae 1991 [where $\cos(i)$ and $d_{5.5}$ are the inclination angle cosine and
distance in 5.5 kpc units respectively]. Differences between these estimates and
previous estimates based on dynamical studies are discussed. It is further shown
that the disk inner radius increases with the high-to-low state transition 
in Nova Muscae 1991. Specifically, our analysis suggests that the inner-disk 
radius
increases to $r_{in}\simeq17$ Scwarzschild radii as the transition to the 
low-hard
state occurs.
\end{abstract}


\section{Introduction}

It has long been known that accreting stellar-mass black holes in 
Galactic binaries
exhibit a "bi-modal" spectral behavior - namely  the  so called high-soft and
low-hard spectral states; see  Liang (1998) for  a recent review.  
High and low in this
context allude to the relative 2-10 keV luminosities in a given system, 
which in turn
responds to the rate at which  mass accretion is taking place. An important 
point is
that the high-soft  state spectra are apparently  a {\it unique} black hole 
signature,
whereas the low-hard state spectral  form has been seen in neutron star binaries
under certain conditions --  typically in low luminosity states of bursters 
(Barret \&
Grindlay 1995). It thus seems reasonable to speculate that this unique spectral
signature is directly tied to the black hole event horizon. This is the
primary motivation for the Bulk Motion Comptonization Model (BMC) introduced
in several previous papers, and recently applied with striking success to a 
substantial
body of observational data. We will not describe the model in detail here, 
but refer
the interested reader to the literature and briefly state its essential 
components:
Compton scattering of thermal X-rays from bulk-motion infall in close 
proximity to
the black-hole event horizon, and a hot Compton cloud obscuring the inner disk
during periods of low-mass-accretion rate 
[Titarchuk, Mastichiadis \& Kylafis 1997,
Titarchuk \& Zannias 1998,  Shrader \& Titarchuk 1998, Laurent \& Titarchuk
1999, Borozdin et al. 1999 (hereafter TMK97, TZ98, ShT98, LT99, BOR99
respectively)].

Given the success of the BMC model in reproducing the high-soft state continuum,
we have begun to investigate potential predictive power. From the spectral shape
and normalization, one can calculate an effective disk radius and the
mass-to-distance  ratio $m/d$. Other quantities  such as  a black hole mass and
mass-accretion rate can be determined if the distance or mass are known
independently.  This is dependent upon an additional factor, $T_h$, 
the so called
``hardening factor''  which represents the ratio of color-to-effective 
temperature. If
one can identify several ``calibrators'',  
i.e. sources for which distance and mass are
well determined,  the model can be applied  to a derive $T_h$ over the available
dynamic range in luminosity. Three such sources are GRO~J1655-40, for which we
have presented extensive analysis in a previous paper, LMC~X-1 for which the
distance is well constrained, and Nova Muscae 1991
(herein NM91), for which the binary  parameters are reasonably constrained. 
In the latter two cases, we have assumed the hardening factor, 
$T_h\simeq2.6$, derived
from our analysis of GRO~J1655-40. The temperature-flux curves we present
provide  a self-consistency test of this assumption.

In this paper, we apply the BMC model to LMC~X-1 and NM91 then, drawing from
previous analysis of GRO~J1655-40, infer some physical parameters of 
these binary
systems based on the model and on methodology detailed in BOR99 and section 3.
In section 4 we offer interpretation of our results including qualitative 
inferences regarding the evolution of the emission-region size over the 
course of the NM91 decline.

\section {PHYSICAL PARAMETER ESTIMATION}

If the physics underlying the emergent spectrum is understood, one can, with
minimal model dependency, use observables to determine parameters of
the system such as the mass-to-distance ratio, the mass accretion rate, and the disk
effective radius. From the inferred color temperature and 
absolute normalization, the
effective area of the emission region can be obtained in terms of the distance 
and $T_h$. The observable surface area can be related to the black hole mass 
using, for example, the Shakura-Sunyaev (1973), herein SS73, 
prescription modified  to treat
electron scattering and free-free absorption and emission  
[Titarchuk 1994 (hereafter Ti94), see also Shimura \& Takahara 1995]. 
The emergent spectrum is then
represented by the integral of a diluted blackbody function characterized by
$T_h^{-4}$.  The flux density $g_{\nu}$, measured in units of
erg~sec$^{-1}$~ster$^{-1}$cm$^{-2}$, is given by:
\begin{equation}
g_{\nu}(E)= C_{N}\cdot 2
\int_{r_{in}}^{\infty}{{E^3(1.6\cdot10^{-9})}\over{\exp[E/T_{max}f(r)]-1}}
rdr
\end{equation}
where $r=R/R_s$ is in Schwarzchild and $m=M/M_{\odot}$ in solar units. The
color temperature distribution $T_{max}\cdot f(r)$ which has the radial
dependence,
\begin{equation}
f(r)={{[1-(r_{in}/r)^{1/2}]^{1/4}(1.36r_{in})^{3/4}}\over{r^{3/4}(1/7)^{1/4}}},
\end{equation}
is normalized to $T_{max}$ at $r=(7/6)^2r_{in}=1.36r_{in}$. The normalization
factor of the equation (2) measured in units phot~sec$^{-1}$~ster$^{-1}$~
cm$^2$ is $C_N=0.91(m/d)^2\cos{i} T_h^{-4}$  where $i$ is the inclination
angle (between  the line of sight and the disk normal), and d is in kpc. We then
apply the mean value theorem to replace the radial temperature distribution with
$T_{col}$ giving
 \begin{equation}
g_{\nu}(E)~=~A_{N}{{E^3(1.6\cdot10^{-9})}\over{\exp[E/T_{col}]-1}},
\end{equation}
\par
\noindent
where $A_{N}= C_{N}r_{eff}^2$. In general, $r_{eff}$ is a function of the
photon energy $E$, however if the energy coverage is such that the ratio
$E/kT_{col}\gtorder2$, $r_{eff}$ is nearly constant and the single temperature
blackbody is a reasonable approximation to the disk spectrum. The
value of $r_{eff}$ then depends only on $T_{col}$ (or $T_{max}$). BOR99
derived $T_h$ using inferred parameters and an averaging method over a
large flux range
[ $(0.3-2.2)\cdot10^8$ ergs s$^{-1}$cm$^{2}$]  obtaining that $T_h\simeq2.6$.
We can then calculate the mass-to-distance ratio
\begin{equation}
\left({{d}\over{m}}\right)^2
={{0.91 T_h^{-4}r_{eff}^2}\over{A_N}}\cos{i}
\end{equation}
and if the distance or mass are known, the mass accretion rate
\begin{equation}
\dot m_{disk} = m{{(T_{max})^4[(7/6)^2r_{in}]^3}\over{(1/7)3^4T_h^4}}.
\end{equation}

\section{VIABILITY OF THE METHOD: APPLICATION TO
 OBSERVATIONAL DATA}

To test the viability of our method, we have collected data for  LMC~X-1, for
which the distance is well constrained by virtue of its location in the LMC - 
herein we adopt the LMC distance of $51.2\pm3.1$~kpc of  Panagia et al (1991) 
as the
distance to LMC~X-1 - and NM91 for which  the binary parameters are
reasonably well constrained and the distance is known to a factor of $\sim2$.
From earlier work on GRO~J1655-40 (BOR99), we determined that
$T_h\simeq2.6$, and demonstrated that the data are strikingly consistent 
with the Temperature-Flux relationship predicted by the SS73/Ti94 model. 
We use this information here.

LMC~X-1, has been observed extensively with various instruments, for example,
HEAO-1 (White \& Marshall, 1984), BBXRT (Schlegel et al. 1994), and RXTE
(Schmidtke et al. 1999). Although luminous, given its $\sim50$kpc distance,
its intensity is much lower than that of typical Galactic XRBs. The impact
on our study is a lack of reliable measurement above $\sim20$~keV.  We found
that typical 2-10~ksec exposures available from the
RXTE data archives did not include reasonable HEXTE (20-200~keV) measurements.
We also examined the publicly available
CGRO/OSSE data. The source appears to be marginally detected in a
200~ksec exposure, a simple power law fit yielding a spectral index consistent
with typical PCA values. However, it is poorly constrained
($\alpha=1.71\pm1.2$), and since the data were not contemporaneously obtained,
we cannot reliably apply our model. Usually, however, the
proportional counter data alone were sufficient to constrain our model
parameters.  NM91 was a bright X-ray transient (also known as GS~1124-683 =
GRS~1121-68) discovered in January 1991 independently by Ginga/ASM and
GRANAT/Watch. It reached a peak intensity of about 6 times the Crab over a
2-10~keV band, and its light curve exhibited the canonical X-ray nova fast-rise
exponential-decay profile  (e.g. Chen, Shrader \& Livio 1997). Subsequent to its
discovery, a number of  follow-up observations with the Ginga LAC detectors
were made, covering portions of the outburst peak and subsequent decay. Both
the high-soft and low-hard states were seen (Ebisawa et al. 1994).

We then followed procedures detailed in BOR99 to
analyze 7 spectra for NM91 and about 30 spectra for LMC-X1. Some
representative results,  for LMC~X-1 (averaged over the sample) are
T=$0.79\pm0.22$, $\alpha=1.63\pm0.24$ and $f=0.17\pm0.23$ (also see Figures
1 and 2.). Cases of particular interest for NM91, are the March 1991 data
($f\sim0$) when the hard-power- law component is not detected,  and September
when the source has entered the low-hard  state 
(T,$\alpha,f$)=($0.29,0.89,0.37$).
A typical peak-outburst high-state spectrum (e.g. 20 January 1991) has
(T,$\alpha,~f$)=($0.81,1.36,0.15$). 
While we were unable to obtain a satisfactory
fit to the March 1991 data, we found that we could model the September 1991
data not only in terms of a thermal Comptonization spectrum (or simple hard
power law), but as a BMC spectrum  with a relatively large $f-$value -- i.e. the
scattered component had become dominant. The disk at this stage is nearly
obscured and our spectral index parameter decreases to values below the
asymptotic limit of the typical high-state case (TZ98).

The apparent broad absorption feature, first noted by Ebisawa et al. (1994), and
modeled in detail by Zycki, Done \& Smith (1998) is evident in our deconvolved
spectrum of 12 January, 1991. However, the depth that we obtain is only about
half that of Ebisawa et al. (1994), 
and it is substantially less pronounced in the 20
January 1991 spectrum.  We do not find strong evidence for such a feature in
subsequent spectra, however, statistics become limiting as the source faded. 
In order to treat the entire evolutionary sequence in a consistent manner, 
we have opted not to incorporate additional features into our analysis.

Following the prescription outlined in the previous section, 
we have constructed a
plot of the empirical flux-temperature relationship for LMC~X-1 and NM91
(Figure 3). The solid curves represent the corresponding relationship in the
context of our disk model for the indicated value of $r_{in}$,
while the dashed-line curves represent the $20\%$ deviation corridor about
the model for $r_{in}=3$. This utilizes the hardening
factor obtained by BOR99.  The theoretical curves
can be reproduced analytically using the following formulae:

\begin{equation}
F=r_{eff}^2T_
{col}^4\int_{4/T_{col}}^{20/T_{col}}{{x^3}\over{e^x-1}}dx.
\end{equation}

We find that $r_{eff}$ can be represented by a fourth-order polynomial of the
form $a_nT_{col}^n$
with the coefficients $a_0=2.674,~4.452,~6.737,~14.33$, $a_1=14.09,~24.57,~43.61,~96.07$,
$a_2=-13.78,~-30.51,~-66.11,~-170.1$, $a_3=14.75,~40.76,~94.68,~282.$,
$a_4=-5.42,~-19.77,~-49.5,~-176.1$ for $r_{in}=3,~5,~8,~17$ 
respectively.
We then applied a normalization factor of $10.1$ to the theoretical values to
match the data.

The data in Figure 3 were obtained by integrating our model flux from
$4-20$~keV. A scale factor of 200 was applied to LMC~X-1 to present
both sources in the same plot (200 is approximately the square of the ratio 
of the mass-to-distance values we derive for NM91and LMC X-1). Most of
the points are associated with the inner disk edge near $3R_s$.
However, as the flux decreases, and NM91 enters the low-hard state, 
{\it the inner
edge recedes outwards to 17 Schwarzchild radii}.  Esin et al.
(1997; hereafter EMCN)  argue that the disk is located near 
the last stable orbit at
$r_{in}=3$ in the high state, consistent with what we infer. However, our
conclusions are incompatible with the EMCN scenario of an accretion flow
extending out to  $r_{in}\sim 10^4$ during decline, 
requiring instead a transition
from the high to low states and optically thin to optically thick structure at
$r_{in}\simeq17$. Zycki et al. (1997) have independently reached the same
conclusion, that the retreat of the inner disk during the high to the low 
state transition
is much slower than predicted by EMCN and optically thick material within $\sim
10-12\, R_s$  is generally present. 
The path we have followed in reaching these
conclusions  however, is distinct from Zycki et al (1997), primarily in 
that those
authors applied an additional model component --  an absorption-edge feature --
from which information  regarding the disk was extracted. We have
applied a self-consistently calculated model without additive components.
Using extensive hydrodynamical and radiative transfer calculations, Chakrabarti
\& Titarchuk (1995) first predicted,  
that the apparent position of the inner edge of
the accretion disk would  be seen as moving to the position of the shock 
location (i.e. between 8 and  15 $R_{s}$). 
Our conclusions here would seem to support that prediction.

\section{DISCUSSION AND CONCLUSIONS}

We have calculated mass-to-distance ratios following the method outlined in the
previous section. For NM91, Orosz et al. (1996) have presented a comprehensive
dynamical study of the quiescent binary system, leading to a maximum probability
contour in $m-\sin(i)$ space with m ranging from about 5 to 7.5 $M_{\sun}$ for
inclination $i\simeq55^o- 65^o$. The most uncertain parameter however is its
distance, with published estimates ranging from 1.4~kpc (Della Valle et al 1991)
to 8~kpc (Cheng et al. 1993). Orosz et al. (1996) who argue that 
$d=5.5\pm1$~kpc based on their characterization of the secondary star, and 
brightness estimates for
the quiescent disk, however, their uncertainty may be understated. We calculate
the BH mass, $m=[24\pm1]\times[1/2/\cos(i)]^{1/2}d_{5.5}(T_h/2.6)^2$ and
mass accretion rate in the disk in Eddington units, 
$\dot m_d \approx 2.5$ [using
formula (5), mass value 24 and $T_{max}\approx 0.95$ keV] where
$d_{5.5}=d/5.5$~kpc. Our mass estimate can be reconciled with Orosz et al.
 (1996)
only if  $d_{5.5}\ltorder0.31$, which would make the outburst luminosity
$L_{3-20}\simeq1.5\times10^{37}$ergs~s$^{-1}$. We note, that for our flux
measurement, if $d_{5.5}=1$ and the mass is near the lower end of the range 
cited by Orosz et al (1996), the luminosity is very nearly 
at the Eddington limit,
whereas in our case it is substantially sub-Eddington. However, 
since stable disk accretion is occurring, super-Eddington levels cannot 
persist and a mass as low as $\simeq5M_{\sun}$ seems unlikely.

For LMC~X-1, we can calculate the mass directly. We find $m=[16\pm
1]\times[(0.5/\cos(i)]^{1/2}(T_h/2.6)^2$  and $\dot m_d\approx 1.7$. 
This large of a mass is consistent with the fact  that LMC~X-1 is relatively 
luminous compared to
typical Galactic X-ray binaries. It would seem, however, to be at odds with
published values obtained from radial velocity measurements of the candidate
optical companion "star 32", of $4-10M_{\odot}$ (Hutchings et al 1987). To
reconcile our result with the upper allowed-range of  their analysis would 
require a low-inclination system, $i\ltorder30^o$, however, this could render 
the ellipsoidal-variability amplitude below the detectability threshold. 
It must be noted
however, that the association of LMC~X-1 with star 32 remains speculative, and
the only significant constraint on the inclination is an apparent lack 
of eclipsing.
We can make an additional argument against the lower end of the Hutchings et al
(1987) mass range: for a given disk inner radius and temperature distribution,
$\dot m$ scales in proportion with $m$. Since $m/d$, and in this case $m$ is an
invariant of our model, for $m\ltorder10$, $\dot m<1$, which is inconsistent 
with a high-soft state source (LT99, BOR99).

The temperature flux relation in NM91 seems to populate two distinct
tracks. This is an indication that the emission area, and 
thus the inner-disk radius, is increasing. 
This is consistent with the previous conclusions of Zycki et al. (1998)
who find $r_{in}\sim13$ -- as illustrated in Figure 3  an inner disk radius of
$r_{in}\simeq17$ is consistent with our analysis.

An issue central to stellar evolution theory is the distribution of black hole 
masses in the Galaxy and contact binaries may provide our only insight.
Dynamical analyses based on optical studies are often
limited due to dust obscuration or the brightness dominance of the disk.
Our method offers, in principle, an independent means of parameter
determination.

\vspace{0.2in}

\centerline{\bf{ ACKNOWLEDGMENTS}}
This work made use of the High-Energy Astrophysics Science Archive Research
Center, the Compton Gamma Ray Observatory and Rossi X-Ray Timing Explorer
science support facilities at the NASA Goddard Space Flight Center. We also wish
to acknowledge Ken Ebisawa for providing the Ginga data and for useful
discussions.

\newpage

\begin{figure}[b!]
\vspace{10pt}
\caption{
 Typical BMC model-fit to the LMC-X1
 high-energy continuum, with the inferred parameters
 in the upper right. Data are from RXTE/PCA observations on
 16 April, 1997. The residuals are normalized to the model values.
 We found the source intensity varied by about a factor of 3 and $T_{COL}$
 ranged from about 0.7 to 0.9 over our data sample.
 }
\end{figure}

\newpage
\begin{figure}[b!]
\vspace{10pt}
\caption{
 Typical BMC model-fit to the NM91
 high-energy continuum.
Model-fit to NM91
 high-energy continuum which yielded the indicated parameters.
 These data resulted from a 20 January, 1991 observation with
 the Ginga/LAC when about 10 days after the initial outburst}
\end{figure}

\newpage
\begin{figure}[b!]
\vspace{10pt}
\caption{
 Temperature-flux relation for NM91 and LMC-X1. The solid
curves represent the model predictions for the indicated inner-disk radii,
calculated according to the method presented in the text.
The dashed curves represent a 20\% deviation about the model.
The points which lie on the lower curve,
corresponding to an inner radius of 17, are from the low-state data
for NM91. }
\end{figure}

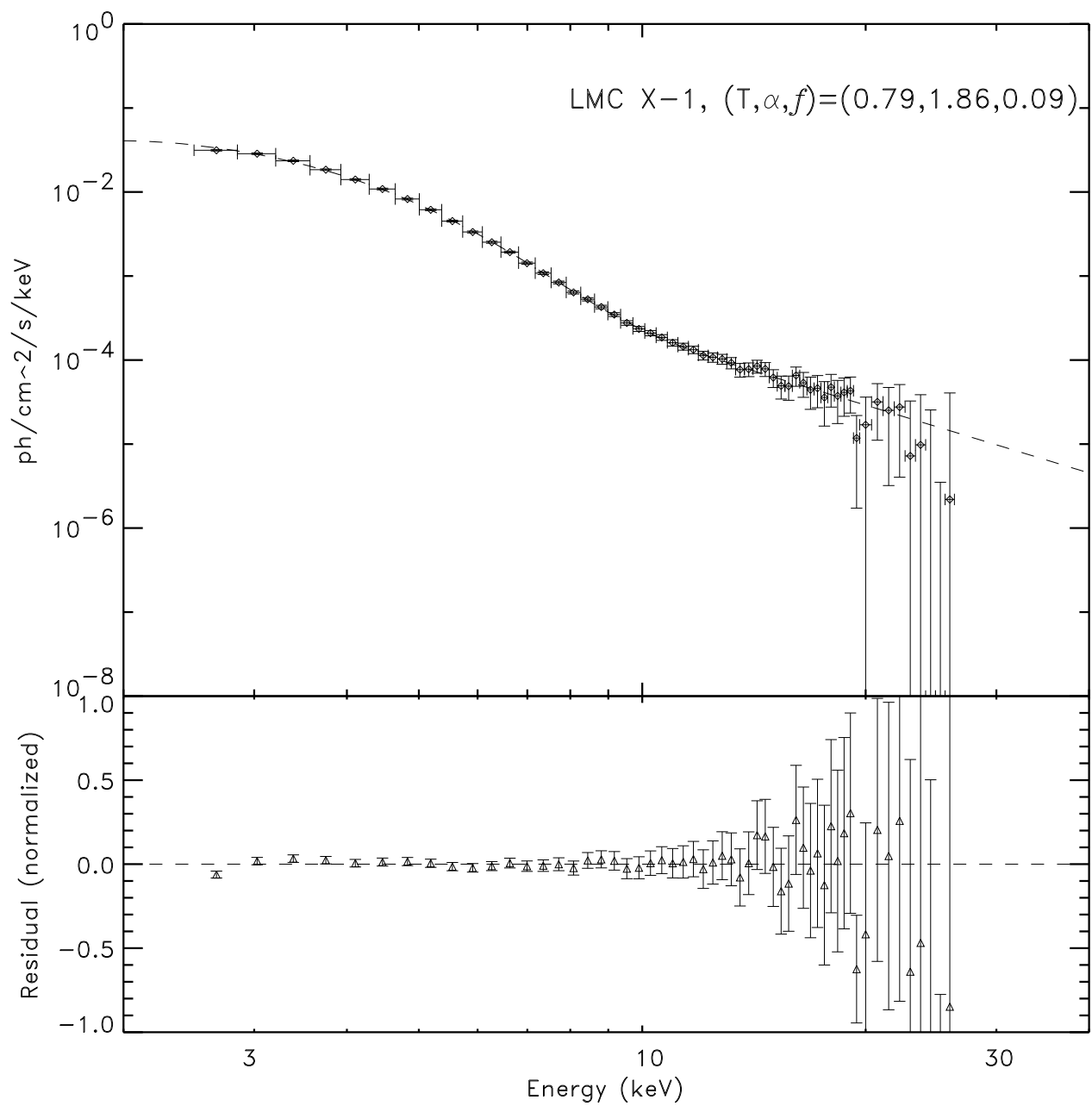
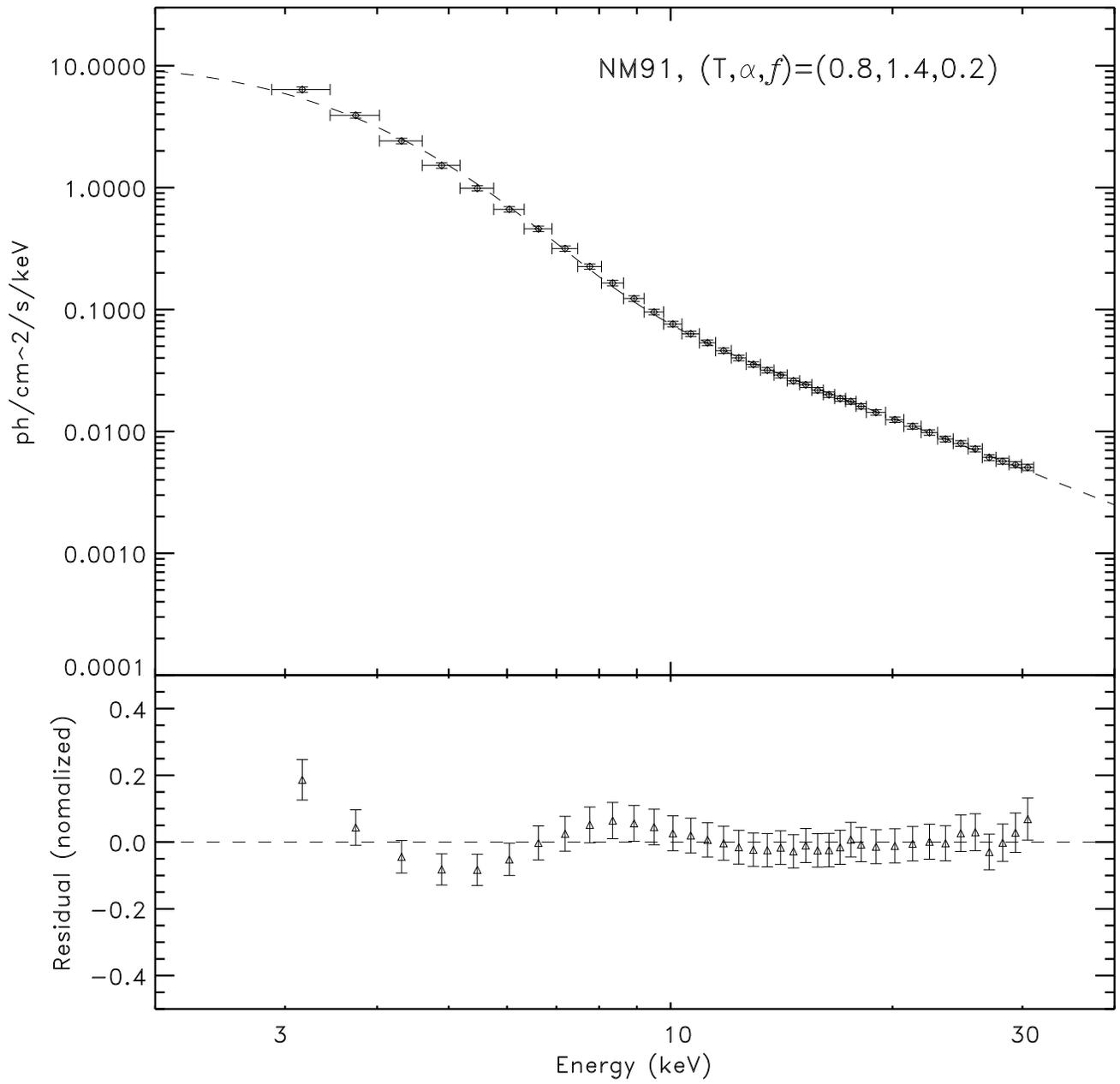
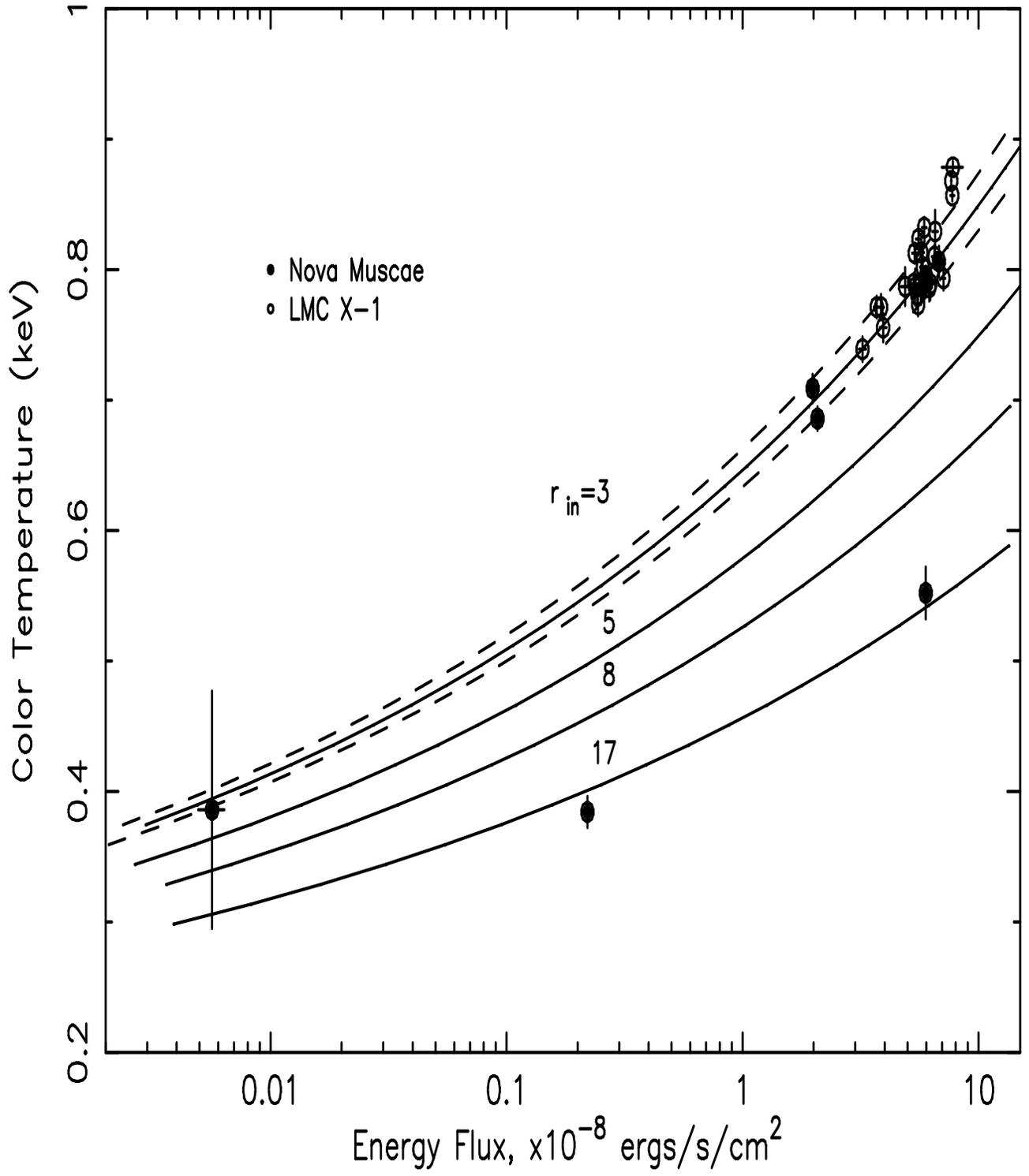
\end{document}